\documentclass[aps,prb,twocolumn,superscriptaddress,longbibliography]{revtex4-2}

\usepackage{graphicx}
\usepackage{siunitx}

\begin{document}

\title{Autonomous Picosecond-Precision Synchronization in Measurement-Device-Independent Quantum Key Distribution}

\author{A.~P.~Pljonkin}
\affiliation{Southern Federal University, Taganrog, Russia}
\email{pljonkin@sfedu.ru}

\begin{abstract}
Measurement-device-independent quantum key distribution (MDI-QKD) eliminates detector side-channel attacks by relocating all measurements to an untrusted intermediate node. However, its practical implementation critically relies on picosecond-level temporal synchronization between spatially separated users. In this work, we present a physically motivated autonomous synchronization algorithm for fiber-based MDI-QKD networks that does not require auxiliary optical channels or shared clock references. The method exploits round-trip optical pulse propagation and statistical signal detection in the presence of Gaussian noise. We derive analytical expressions for false-alarm probabilities, quantify detection reliability, and demonstrate through numerical modeling that synchronization accuracy better than 10~ps is achievable for channel lengths up to 100~km with realistic optical power levels. The proposed approach improves the scalability and robustness of MDI-QKD architectures and is directly applicable to metropolitan and backbone quantum networks.
\end{abstract}

\maketitle

\section{Introduction}

Quantum key distribution (QKD) provides information-theoretic security based on fundamental quantum-mechanical principles. Among existing protocols, measurement-device-independent QKD (MDI-QKD) has emerged as a particularly robust architecture, as it closes all detector side channels by design. In MDI-QKD, both users transmit quantum states to an intermediate node that performs Bell-state measurements, which may be fully untrusted without compromising security\cite{Kulik2025,Lo2012}.

Despite this conceptual advantage, MDI-QKD imposes exceptionally stringent technical requirements on temporal and phase synchronization. Successful two-photon interference at the central beam splitter requires that optical pulses generated by independent laser sources arrive within a narrow temporal window, typically on the order of several picoseconds. Any residual timing mismatch directly reduces Hong-Ou-Mandel interference visibility and degrades the secure key rate.

In deployed fiber networks, synchronization is challenged by environmental perturbations such as temperature-induced fiber length fluctuations, acoustic vibrations, and slow mechanical drift. Existing synchronization techniques often rely on additional optical fibers, high-power reference signals, or centralized clock distribution, which limit scalability and practical deployment. These limitations motivate the development of autonomous synchronization methods that are physically robust and compatible with untrusted network nodes.

\section{MDI-QKD Synchronization Constraints}

In a standard fiber-based MDI-QKD configuration, Alice and Bob independently generate weak coherent pulses and transmit them to a central node Charlie. Charlie performs partial Bell-state measurements using a beam splitter and single-photon detectors. The probability of successful interference depends on temporal overlap, spectral indistinguishability, and phase coherence of the incoming pulses.

The MDI-QKD protocol was created as a response to vulnerabilities in measurement components in quantum cryptography \cite{Ponosova2025}. Traditional QKD systems are susceptible to detector attacks; however, in the new architecture, measurements are placed in a separate node, allowing communication channel security to be maintained even if the measurement equipment is compromised. The system involves three main components: two legitimate users (Alice and Bob), an intermediate node (Charlie), and a classical communication channel. This organization creates a structure where critically important measurements are performed in a potentially insecure environment without compromising the confidentiality of transmitted data \cite{Xu2013}.

Legitimate users independently generate sequences of coherent states with different intensity levels. Information transmission is performed by encoding in randomly chosen bases. A feature of the MDI protocol is the use of three types of states with different intensities, providing resistance to attacks using fake states \cite{Tang2014}. The prepared states are transmitted via quantum channels to the intermediate node, where photon interference occurs. A key requirement is strict synchronization of photon arrival from both users, ensuring correctness of the interference pattern \cite{Comandar2015}. The intermediate node performs measurements in the Bell basis using single-photon detectors and a time-coincidence registration system. Measurement results, namely detector coincidence events, are transmitted via the public channel without disclosing which specific detectors registered photons \cite{Yin2016}. At the final stage, users select events where both transmitted signal states, compare used bases via the public channel, and form a raw key from events with matching bases. The final stage is error correction and privacy amplification, yielding the final secret key \cite{Zhou2017}.

The problem of ensuring precise temporal and phase synchronization remains an important technical task in building practical quantum key distribution systems with measurement-device independence \cite{Chen2023}. Analysis of synchronization methods shows that accuracy requirements in MDI-QKD significantly exceed similar parameters of classical quantum communication systems. Modern approaches to temporal synchronization demonstrate evolution from classical reference pulse transmission schemes to hybrid solutions. Research shows that combining wavelength division multiplexing (WDM) with adaptive filtering algorithms allows achieving synchronization accuracy up to \SI{10}{ps} in metropolitan networks up to \SI{50}{km} long \cite{Wang2024}. At the same time, an acceptable level of quantum noise is maintained, confirmed by experimental data obtained in real operating conditions. In the field of phase synchronization, a transition to systems with predictive drift compensation is observed. Works \cite{Kumar2023, Rodriguez2024} demonstrate the effectiveness of using machine learning to predict temperature changes in fiber-optic lines. Quantum synchronization methods are considered a promising direction. Experimental studies \cite{Schroder2023} confirm the possibility of using temporal correlations in entangled photon pairs for synchronization with accuracy up to \SI{100}{fs}. However, practical implementation of such systems requires overcoming serious technological barriers related to the complexity of generating and detecting entangled states in field conditions.

The most common method for time coordination of nodes in fiber-optic MDI-QKD systems is using a single master frequency generator for all participants. In this configuration, the generator is usually placed at Charlie’s device and connected by electrical cables to control modules of Alice and Bob. This synchronization method has proven effective in laboratory conditions for testing system principles. However, a key drawback of this approach is that Alice and Bob devices cannot be placed far from Charlie. Thus, although the method is ideal for testing, it is unsuitable for commercial deployment. The second method is based on calculating cross-correlation between sent and received qubit sequences \cite{Cochran2021}. Its weak point is the requirement to use single-photon detectors --- the most technologically complex element in modern quantum key distribution systems. The third approach involves using powerful optical pulses transmitted via the quantum channel or a separate fiber. Using an additional fiber complicates system architecture. Patent \cite{Patent1} describes a system and method for compensating time shift of optical pulses in quantum channels of a quantum key distribution device with an untrusted central node. The system proposed by the patent authors is based on using an adjustable optical delay line in one of the untrusted node inputs, included in a feedback loop with single-photon detectors, allowing control of laser pulse overlap at the beam splitter of the untrusted node with accuracy up to units of picoseconds. It is assumed that this solution should improve interference in real operating conditions, thereby increasing maximum distance and secret key generation rate. Patent \cite{Patent2} describes a quantum key distribution system with an untrusted central node and a method for preparing quantum states for the quantum key distribution protocol with an untrusted central node. The work presents an approach using passive phase randomization (i.e., without a phase modulator) while ensuring high interference quality. A theoretical description of the proposed method and experimental comparison with the existing method based on a continuous-wave laser are presented. Patent \cite{Patent3} describes a synchronization device for quantum key distribution systems operating on side frequencies in a star topology. Synchronization here uses a separate optical fiber, a pulsed coherent distributed feedback laser source, and a beam splitter. Interesting research is described in article \cite{Rudavin2022}, proposing a method for automatically adjusting Alice and Bob frequencies to Charlie’s frequency. The work demonstrates a frequency synchronization protocol for MDI-QKD setups using a laser and a 50/50 beam splitter for synchronization signal transmission, as well as a phase-locked loop for computing the control signal value.

\section{Autonomous Synchronization Algorithm}

The proposed synchronization algorithm adapts concepts from autocompensating QKD systems to the MDI-QKD topology. Each user station is equipped with an optical circulator and a classical photodetector, while the untrusted node contains weak tap couplers and reflective elements. No trusted hardware or timing reference is required at the central node.

Each station independently performs a round-trip time-of-flight measurement by transmitting optical pulses toward Charlie and detecting the reflected signal. The detection process is formulated as a statistical hypothesis test in discrete temporal windows. Importantly, the sequential search terminates upon first signal detection, minimizing synchronization latency.

After both users determine their respective propagation delays, Charlie compares the measured arrival times and computes a relative temporal offset. One station then applies a compensating delay by adjusting the emission timing of its laser pulses. Fine alignment is achieved by monitoring two-photon interference visibility.

The synchronization process in quantum key distribution systems is a set of procedures aimed at establishing and maintaining temporal correspondence between distributed system nodes. Its fundamental task is to ensure coherent interaction of single-photon signals propagating along independent optical paths. Without precise synchronization, correct interference of photon states, which is the basis of the MDI-QKD protocol, becomes impossible. From the perspective of optical signal detection, synchronization solves the task of identifying time windows in which single-photon states are expected to arrive. Signal search is performed through analysis of temporal correlations between detected events and reference clock sequences.

For MDI-QKD, maintaining synchronization under conditions of phase drifts and temperature variations affecting optical signal propagation paths is particularly challenging. Stabilization of relative phase shifts requires creating feedback systems continuously compensating changes in path difference between interfering signals. In systems with long fiber-optic lines, synchronization is additionally complicated due to group delay instabilities caused by mechanical stress and temperature gradients. Overcoming these limitations requires developing specialized time coordination protocols adapted to real telecommunication channel conditions. Successful implementation of such protocols enables building scalable quantum networks with guaranteed security parameters.

We modified the synchronization algorithm used in two-pass auto-compensation QKD systems \cite{Pljonkin2017} for MDI-QKD topology and made minor changes to station design. A feature of the algorithm is its autonomy, i.e., the ability to perform synchronization without coordinating data via a public (service) communication channel between stations. Figures 1 and 2 show schematic diagrams of optical parts of distributed QKD systems using polarization and phase encoding. Optical circulators and photodetectors are added to Alice and Bob station designs. Beam splitters are added to Charlie station’s input path, diverting part of radiation to reflective mirrors.

\begin{figure}[htbp]
    \centering
    \includegraphics[width=0.9\linewidth]{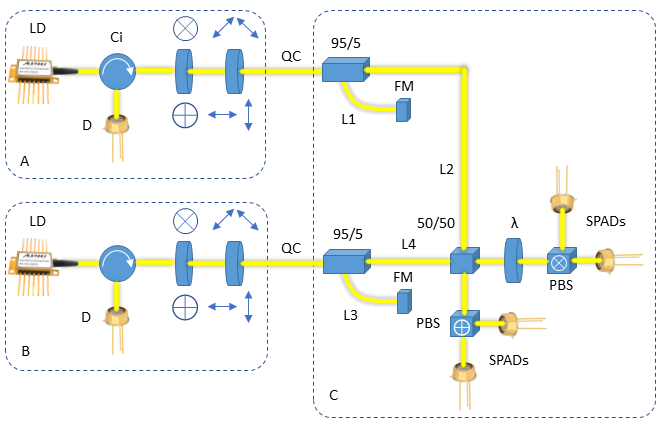}
    \caption{Optical part of MDI-QKD scheme with polarization encoding.}
    \label{fig:polarization}
\end{figure}

Figure 2 shows the optical path diagram of MDI QKD system using phase encoding of photon states.

\begin{figure}[htbp]
    \centering
    \includegraphics[width=0.9\linewidth]{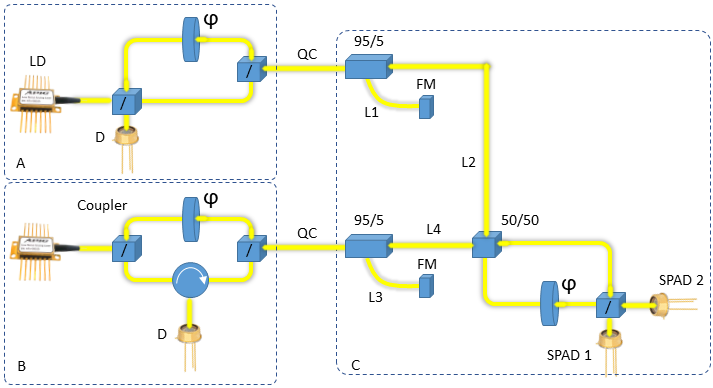}
    \caption{Optical part of MDI-QKD scheme with phase encoding. LD --- radiation source; D --- photodetector; Coupler --- optical beam splitter; SPAD --- avalanche photodetectors; 50/50 --- X-type beam splitter; 95/5 --- Y-type beam splitter; FM --- mirror for reflecting optical signal; PBS --- polarizing beam splitter; Ci --- optical circulator; L --- optical arms; QC --- quantum communication channel; A --- Alice station; B --- Bob station; C --- Charlie station (untrusted node).}
    \label{fig:phase}
\end{figure}

Beam splitters are also integrated into Charlie node design and optical circulators with photodetectors in Alice and Bob stations. For phase encoding, the circulator can be placed directly in one interferometer arm or the photodetector can be connected to one interferometer beam splitter input. When diverting optical radiation from one interferometer arm, we unambiguously fix the pulse at the photodetector. The situation changes when connecting the photodetector to one beam splitter output. Here detection will depend on the interference pattern (outcome). Alternatively, an optical circulator can be installed between the interferometer and radiation source (not shown in figure). In this case, signal detection will occur unambiguously.

Optical signal with wavelength \SI{1550}{nm} enters the Mach-Zehnder interferometer beam splitter of Bob station, where it splits into two equal power pulses. Following different arms, pulses enter the quantum channel with time delay. At Charlie station input, pulses enter a beam splitter where a smaller part of energy is diverted to an optical mirror and reflected. During backward propagation, the signal enters the optical circulator in the interferometer arm (Figure 2) and is registered by the photodetector. Charlie station electronics must know when to activate avalanche photodetectors to record the interference result of signals arriving at the beam splitter. Alice and Bob stations must know the exact distance to the main beam splitter in Charlie and exact pulse generation time for each station. This is required so that single photons from Alice and Bob arrive at Charlie’s beam splitter simultaneously. Accuracy (error) of photon arrival at the beam splitter is about \SI{10}{ps}. Since stations are located at different distances from Charlie, we propose performing time correction (delay) of signals by adjusting radiation sources. Charlie electronics knows exact lengths of its interferometer arms and thus can activate SPDs at the right moment with required time delay relative to signal arrival at the beam splitter. This delay value will be constant for a specific Charlie implementation. Furthermore, optical path lengths from 95/5 beam splitter to interferometer beam splitter and from 95/5 beam splitter to reflecting mirror must be equal (L1, L2 and L3, L4). Knowing optical pulse travel time from radiation source to reflecting mirror and back to photodetector, Alice and Bob stations can independently determine distances to Charlie’s beam splitter. Note that as 95/5 beam splitter, an electronically controlled version can be used that disables the necessary optical signal output (e.g., disable output to reflecting mirror during quantum protocol operation).

Consider the synchronization process scheme through a sequential algorithm of events (Figure 3).

\begin{figure}[htbp]
    \centering
    \includegraphics[width=0.9\linewidth]{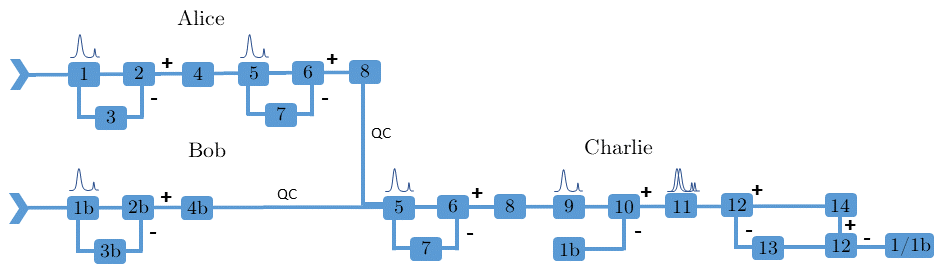}
    \caption{Synchronization process of stations in MDI-QKD configuration. Symbols + and – mean positive and negative condition (event) result.}
    \label{fig:process}
\end{figure}

The entire process can be divided into several stages: Alice measures quantum channel length; Bob measures quantum channel length; Alice informs Charlie of obtained data; Charlie performs gating and adjustment based on Alice’s data; Bob informs Charlie of obtained data; Charlie performs gating and adjustment based on Bob’s data; Charlie computes delta between signals from Alice and Bob, informs one station; All participants perform precise time adjustment.

Blocks 1--3 in Figure 3 reflect the process of measuring distance to Charlie’s beam splitter. Distance is expressed through time with accuracy up to \SI{10}{ps}. This process is implemented through sequential analysis of time intervals with stepwise increase of detection time delay for Alice station photodetector. 1 --- pulse sending; 2 --- detection result in specific time interval; 3 --- increase of detection time delay. After determining signal time window, Alice transmits data about pulse repetition period to Charlie (4) and sends command to activate SPD (5). Charlie electronics now knows pulse repetition period from Alice and, receiving command, activates photodetector. The first registered signal from Alice starts Charlie’s cyclic adjustment (5--7). Such adjustment is implemented through sequential analysis of adjacent time intervals relative to known gating time. The algorithm assumes analysis of intervals of \SI{10}{ps} duration with time shift ±\SI{2}{ns}. Operation result will be found interval with signal pulse (8). For unambiguous selection of one narrow time interval, multiple checking of each time window is used. Success of stage (block 8) means that Alice knows exact optical pulse travel time to Charlie’s beam splitter, and Charlie possesses information about exact arrival time of Alice’s pulses at SPD. We can say that at this intermediate stage Alice and Charlie are synchronized.

By this time, Bob station has already passed stages 1--4, i.e., possesses information about pulse propagation time to Charlie’s beam splitter. Now stages 5--8 repeat for Bob station. Alice station does not stop. Electronically controlled beam splitter blocks pulses arriving from Alice to Charlie station. At stage 9, Charlie, knowing arrival times of pulses from Alice and Bob, computes time delta (Figure 5) and transmits its value to Bob for correction (introducing time delay). Block 10 compares delta, i.e., whether pulses from Alice and Bob arrive simultaneously. At this stage, Charlie electronics controls beam splitter outputs, switching them alternately. The latter is done for more precise adjustment. After coordination, at stage 11 both beam splitter outputs open and pulses from Alice and Bob enter Charlie station. Then at 12 interference presence is recorded. If interference is present, synchronization is successful. If absent, Charlie performs fine adjustment within ±\SI{60}{ps}. If result is negative, calibration steps are repeated (1/1b).

Consider physical and mathematical models of distance measurement process to Charlie’s beam splitter from one station (blocks 1--3 in Figure 3). Optical path from station, e.g., Bob to Charlie’s beam splitter (and hence to SPD) and back structurally corresponds to auto-compensation quantum key distribution system scheme. In such systems, the task of optical signal search is solved by a known method \cite{Galliardi1978}. Our works provide detailed description of synchronization algorithm for such QKD systems and propose various improvement options \cite{Pljonkin2024}.

Pulse repetition period $T$ is known, chosen based on maximum possible optical path length to avoid reflected signal overlap. Pulse duration $\tau = \SI{1}{ns}$, time window duration $\tau_{w} = \SI{2}{ns}$. Number of time windows accordingly equals $N = \frac{T}{\tau_{w}}.$ Since classical photodetectors are used in Alice and Bob stations, signal is described by normal distribution (Gaussian distribution). To decide signal presence, threshold $P$ is set for photodetector amplitude. If measured signal exceeds this threshold, decision is made about its presence. Reflected signal search is performed by sequential analysis of all $\tau_{w}$. In idealized version, algorithm provides continuous sequential analysis of all time windows, but in real systems photodetector requires recovery time up to \SI{100}{ns}, so after each pulse sending one time window is analyzed. Main source of false alarms in this case is photodetector dark current and amplification noise (e.g., shot noise, thermal noise). These noises are random and in our case also described by normal distribution. Probability that noise sample exceeds trigger threshold ($q$) in one time window is computed via Q-function \cite{Gao2007} and considers noise standard deviation $\sigma$:

$$P_{\text{false}} =\left(\frac{q}{\sigma}\right)$$

Since we have $N$ independently polled time windows, probability that no false alarm occurs in any of them equals product of probabilities for each window:

$$P = {(1 - P_{\text{false}})}^{N}$$

Consequently, probability that at least one false alarm occurs per period equals

$$P_t = 1 - {(1 - P_{\text{false}})}^{N}$$

Formula shows that $P_t$ exponentially depends on ratio $\frac{q}{\sigma}$. Even slight threshold increase sharply reduces false alarm probability. Unlike signal detection algorithm used in auto-compensation QKD systems, we propose not analyzing all time windows. The latter means sequential analysis of time windows stops after signal detection (Figure 4).

\begin{figure}[htbp]
    \centering
    \includegraphics[width=0.9\linewidth]{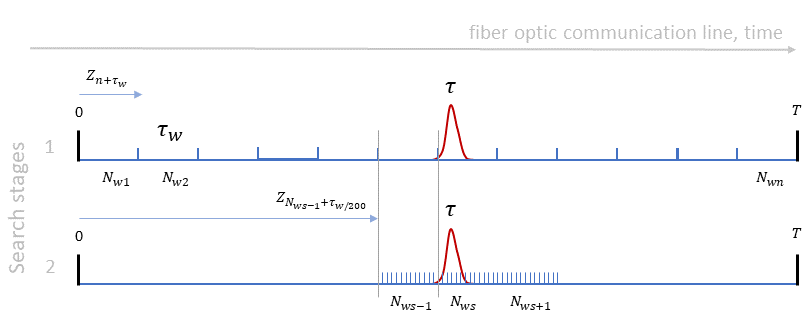}
    \caption{Signal time window search}
    \label{fig:search}
\end{figure}

At first search stage after each pulse sending $\tau$, one time window $N_{w}$ is analyzed. Time delay $Z$ sequentially increases by value $\tau_{w}.$ After detecting signal time window $N_{ws}$, search stops and algorithm proceeds to second (refinement) stage. Here time intervals $N_{ws - 1},\ N_{ws}\ \text{and}\ N_{ws + 1}$ are split into subintervals of duration $\tau_{w}/200$, i.e. \SI{10}{ps}. Error margin in $N_{ws + - 1}$ is taken to account possible signal shift due to, e.g., external destabilizing factors affecting optical fiber. Operation result for blocks 1/1b -- 3/3b (Figure 3) will be knowledge of pulse travel time to Charlie’s beam splitter with accuracy up to \SI{10}{ps}.

Blocks 5--7 are responsible for adjusting Charlie’s SPD after one station sends data about pulse repetition period. After Charlie’s photodetector is ready to receive pulses, one station activates laser diode and first received signal starts cyclic adjustment process. Adjustment algorithm is similar to procedure of stage 2 in Figure 4. Gating pulse is located in time window $N_{ws}$, and analysis process begins with stepwise scanning of each interval of duration $\tau_{w}/200$.

\begin{figure}[htbp]
    \centering
    \includegraphics[width=0.9\linewidth]{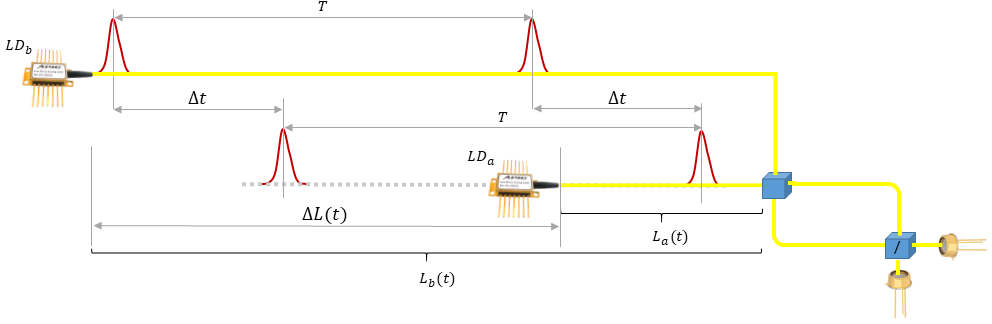}
    \caption{Optical signal desynchronization}
    \label{fig:desync}
\end{figure}

Figure 5 shows time diagram of pulses arriving at Charlie’s beam splitter (photodetector) at different times. Naturally, time difference with same repetition period is caused by different quantum channel lengths. Computing time delay between pulses, Charlie sends data $\Delta L(t)$ to one station. Technically, compensation delay is performed by shifting laser diode activation time. Operation repeats until synchronization of two signals with accuracy up to \SI{10}{ps} is achieved. Note that for two signals condition of equality of repetition period must be satisfied. Phase adjustment for obtaining interference is performed by similar method (Figure 4), but with time error of \SI{60}{ps}.

\section{Statistical Detection Model}

Let’s simulate proposed algorithm for detecting signal time window considering above initial data. Note that at first stage sequential analysis of time windows stops upon signal detection. At second stage all time subintervals are analyzed. Analyze graphs of dependencies of correct signal time window detection probability on radiation power considering different fiber lengths (Figure 6).

\begin{figure}[htbp]
    \centering
    \includegraphics[width=0.9\linewidth]{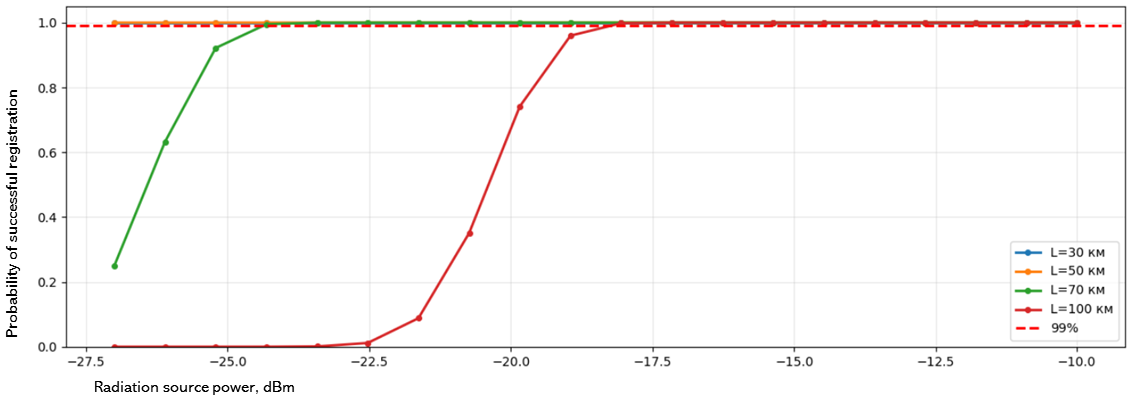}
    \caption{Influence of radiation power on detection probability}
    \label{fig:power}
\end{figure}

Simulation results show that to achieve probability above 99\% at fiber length \SI{100}{km}, sufficient condition is using optical emitter power of \SI{-17.7}{dBm}. If considering quantum channel length less than \SI{50}{km}, radiation power of \SI{-10}{dBm} will provide 100\% detection probability. Simulation considered photodetector quantum efficiency of 20\%, fiber attenuation of \SI{0.2}{dB/km} and losses of \SI{0.3}{dB} at each connector adapter.

Real QKD systems use radiation sources that in synchronization mode have power ranging from \SI{-5}{dBm} to \SI{-19}{dBm}. Figure 7 presents results of analysis to determine optimal sample size in each time interval at first and second stages at fixed laser power $P = \SI{-15}{dBm}$ to achieve correct detection probability >99\%.

\begin{figure}[htbp]
    \centering
    \includegraphics[width=0.9\linewidth]{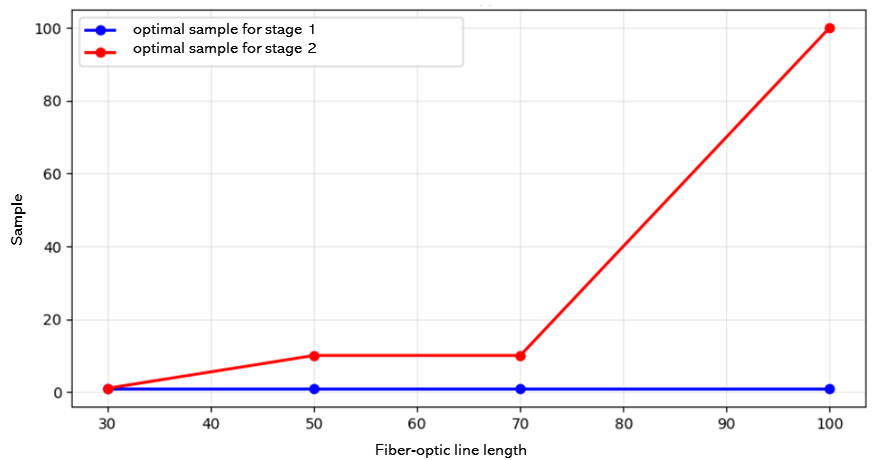}
    \caption{Dependence of sample size on fiber length}
    \label{fig:sample}
\end{figure}

It can be seen that at first stage sufficient condition is single analysis of time window for all fiber length values. When moving to second stage, sample size needs clarification. Thus to account for maximum possible distance of \SI{100}{km}, each subinterval must be polled at least 100 times.

The detected signal amplitude is modeled as a Gaussian random variable, reflecting dominant contributions from thermal and shot noise. A fixed detection threshold is applied, and the probability of false detection in a single temporal window is given by
\begin{equation}
P_{\mathrm{FA}} = Q\!\left(\frac{A_{\mathrm{th}}}{\sigma}\right),
\end{equation}
where $A_{\mathrm{th}}$ is the threshold amplitude, $\sigma$ is the noise standard deviation, and $Q(\cdot)$ denotes the complementary error function.

For $N$ independently scanned windows, the probability of at least one false alarm scales as $1-(1-P_{\mathrm{FA}})^N$. This dependence motivates early termination of the search once a valid signal is detected.

\section{Numerical Results}

Monte Carlo simulations were performed assuming a detector quantum efficiency of 20\%, fiber attenuation of 0.2~dB/km, and connector losses of 0.3~dB. For a 100~km fiber link, a launch power of $-17.7$~dBm yields a detection probability exceeding 99\%. For links shorter than 50~km, near-unity detection probability is achieved at $-10$~dBm.

The refinement stage requires increased sampling density for longer channels. At 100~km, reliable localization of a 10~ps subinterval requires approximately 100 detection events per subwindow.

\section{Discussion and Conclusions}

We have presented an autonomous synchronization algorithm for MDI-QKD systems that achieves picosecond-level precision without auxiliary channels or trusted central hardware. The method is grounded in a physically transparent statistical detection framework and is compatible with realistic fiber-network conditions.

The proposed approach enhances scalability of MDI-QKD networks and reduces architectural complexity. Future work will address experimental validation and analyze security implications related to authentication during the synchronization phase.

\begin{acknowledgments}
This work was supported by the Russian Science Foundation under Grant No.~25-29-00007.
\end{acknowledgments}

\bibliography{references}

\end{document}